\documentclass[aps,twocolumn]{revtex4}
\usepackage{graphicx,subfigure}
\usepackage{amsmath,amssymb}
\usepackage{bm}
\newcommand{\Dslash}{D\hspace{-1.6ex}/\hspace{0.6ex} }
\newcommand{\be}{\begin{eqnarray}}
\newcommand{\ee}{\end{eqnarray}}

\def\slashchar#1{\setbox0=\hbox{$#1$}           
   \dimen0=\wd0                                 
   \setbox1=\hbox{/} \dimen1=\wd1               
  \ifdim\dimen0>\dimen1                        
 \rlap{\hbox to \dimen0{\hfil/\hfil}}      
  #1                                        
 \else                                        
    \rlap{\hbox to \dimen1{\hfil$#1$\hfil}}   
    /                                         
 \fi}                                         %

\begin{document}

\title{Instanton-dyon Ensemble with two Dynamical Quarks: the Chiral Symmetry Breaking}

\author{ Rasmus Larsen   and Edward  Shuryak }

\affiliation{Department of Physics and Astronomy, Stony Brook University,
Stony Brook NY 11794-3800, USA}

\begin{abstract}
This is the second paper of the series aimed at understanding the ensemble of instanton-dyons,
now with two flavors of light dynamical quarks. The partition function is appended by the fermionic factor,
$(det T)^{N_f}$ and Dirac eigenvalue spectra at small values are derived from the numerical simulation
of 64 and 128 dyons. Those spectra show clear chiral symmetry breaking pattern at high dyon density.

\end{abstract}
\maketitle

\section{Introduction}
\subsection{Instanton-dyons and confinement}
At high temperatures QCD matter is in form of quark-gluon plasma (QGP) state, which is weakly coupled
because of the asymptotic freedom phenomenon. The topological solitons to be discussed below
have large action $S=O(1/\alpha_s)\gg1 $ and are therefore strongly suppressed, $\sim exp(-S)$.
However, as $T$ decreases toward the deconfinement transition, the coupling grows and such objects become 
important.

 The non-trivial configurations of interest are Instantons \citep{Belavin:1975fg},
 which in the Euclidean finite temperature formulation are known as
  Calorons. Such solutions have been generalized to the case of non-zero expectation value of the Polyakov loop
by  Lee-Li-Kraan-van Baal   in refs \cite{Kraan:1998sn,Lee:1998bb} and are known as LLKvB calorons.
  An important novel feature of these solution was the realization of instanton substructure: 
  each LLKvB caloron 
consists of $N_c$ objects, known as instanton-dyons (or instanton-monopoles).

Color confinement phenomenon has many manifestations, and thus many definitions. In this series of papers
we focus on one particular aspect of it, namely on the shift of the vacuum expectation value of the Polyakov loop
from its ``trivial value" $<P>\approx 1$ at high $T$ to small  $<P>\approx 0$ at $T < T_c$. 
Multiple numerical simulations in the framework of lattice gauge theory have documented such shift,
as well as modification of the effective potential $V(P,T)$ with $T$ leading to it. Since a contribution of the quarks (and non-diagonal gluons) to thermodynamical quantities is proportional to (powers) of $<P>$, vanishing of it,
effectively switches off quark-gluon plasma contributions. So, 
in papers of this series we  focus on the calculation of this effective potential
and on the deconfinement phase transition phenomenon.

  
  Another manifestation of confinement  is a disordering of large Wilson loops. It has been argued in  \cite{Gerhold:2006sk} that an ensemble of instanton-
 dyons can generate the expected area law. However, this issue is rather subtle and depends on the
 infrared tails of the soliton fields, which are modified by screening effects and thus are not robust
 enough to be conclusive. 
 One more
  approach to the confinement issue is reached via the static quark potentials, which do exist
 at any $T$ and were extensively studied on the lattice. We intend to calculate those in our approach later.
 Finally, a classic formulation of confinement includes absence of color degrees of freedom from vacuum spectra, at $T=0$. Addressing it directly is not possible for the type of models we discuss, since the calorons and instanton-dyons themselves become difficult to use at sufficiently low $T$.
 
 The idea that the effective potential of the Polyakov loop $P$ is due to the back reaction of the instanton-dyons 
 goes back to Diakonov and collaborators \cite{Diakonov:2009ln}, who provided the first estimates indicating how this may happen, but were unable to prove it.  Using the so called ``double-trace
deformation of Yang-Mills theory", at large $N$  on $S^1\times R^3$,
 Unsal and Yaffe~\cite{Unsal:2008ch}  argued  that 
 there can be confining behavior, with unbroken center symmetry, even in weak coupling. 
This construction was extended  by Unsal and collaborators \cite{Poppitz:2011wy,Poppitz:2012sw,Poppitz:2012nz}
 to a class of  deformed supersymmetric theories with soft supersymmetry breaking. In such a
 setting , with  weak coupling and
 an exponentially  $small$ density of the dyons, the minimum of the potential is at the confining 
  value of $P$  induced by the repulsive interaction in the dyon-anti-dyon pairs (called  
 $bions$ by the authors). (The supersymmetry was needed to cancel the perturbative Gross-Pisarski-Yaffe-Weiss (GPYW)  holonomy potential , which otherwise favors trivial value $<P>=1$.
Sulejmanpasic and one of us \cite{Shuryak:2013tka} have proposed a simple model, with ``repulsive cores" in the dyon-antidyon channel,
which can generate confining $V(P)$ at certain temperature $T_c$ in pure gauge theory. 
 
To evaluate the free energy of the instanton-dyon ensemble we
 performed numerical simulations for pure gauge $SU(2)$ theory,  in the first paper of this series \cite{Larsen:2015vaa},
to be below referred as I. The essential element was inclusion of dyon-antidyon
interactions, determined in the previous work \cite{Larsen:2014yya} using  a gradient flow method.
A similar conclusion has been recently reached by Liu, Shuryak and Zahed 
 \cite{Liu:2015ufa} using  analytic mean field theory. It however uses the mean field approximation
 which is only applicable
 for  high enough dyon density, or $T<T_c$.

\subsection{Quarks in the instanton-dyon ensemble}

 
 In 
 this paper we include quarks,  fermions in the fundamental color representation, to the instanton-dyon ensemble.
 Those will be 
 referred  to as ``dynamical quarks", since the so called fermionic determinant will be 
 included in the ensemble measure.
 
The topological objects, instantons and  instanton-dyons,  have a certain number of 4-dimensional
zero modes prescribed by
the topological index theorems. Topology ensures that any smooth deformation of the objects themselves does not 
shift fermionic eigenvalues from zero.

When the ensemble of topological solitons is dense enough, the fermionic zero modes can
  collectivize and produce the so called Zero Mode Zone (ZMZ).
For an ensemble of instantons this phenomenon has been studied in
  great detail in the 1980's and 1990's, for a review see \cite{Schafer:1996wv}.
 The main physical phenomenon associated with ZMZ is the spontaneous breaking of the $SU(N_f)$
 chiral symmetry,  ``chiral breaking" for short.

\begin{table}[h]
\begin{tabular}{ l | c | c | c | r  }
  \hline  
 &  $M$ & $\bar{M}$ & $L$ & $\bar{L}$ \\   
   \hline                     
e & 1 & 1 & -1 & -1 \\
m & 1 & -1 & -1 & 1 \\
  \hline  
\end{tabular}
\caption{Quantum numbers of the four different  kinds of the instanton-dyons of the SU(2) gauge theory.
The two rows are electric and magnetic charges.
}
\label{tab1}
\end{table} 

   In the case of $SU(2)$ gauge group there are only two types of instanton-dyons, called $M$ and $L$ types
   (also known as BPS and ``twisted" or KK ones), their electric and magnetic charges are given in Table \ref{tab1} .
Physical (antiperiodic in time direction) fermions have zero modes on the $L$  dyons.  
The zero modes produce the simplest effect 
 of the dynamical fermions - binding of the $\bar L L$ dyon pairs into ``molecules" , studied
  by  Shuryak and  Sulejmanpasic
\cite{Shuryak:2012aa}. The first numerical simulations with fermions were done by 
Faccioli and Shuryak \cite{Faccioli:2013ja}, who studied 1, 2 and 4 flavor theory with the SU(2) color:
they found chiral symmetry breaking in the first two cases, but  the last one, $N_f=4$ appeared marginal.
Many technical aspects of our work follows their setting.

Recent work by   Liu, Shuryak and Zahed \cite{Liu:2015jsa} was also devoted to the role of quarks
in the dense confining instanton-dyon ensemble. Their basic conclusion is that in this regime
the quark condensate, signaling chiral symmetry breaking, satisfies certain 
universal gap equation, which has non-zero solutions provided the number of quark flavors $N_f<2 N_c$. 
So, the border case for 2 colors is $N_f=4$, which is also a near-critical  one according to Ref. \cite{Faccioli:2013ja}.

In the present work we focus on the simplest case with the spontaneous breaking of chiral symmetry, with only
two quark flavors $N_f=2$. The central issue addressed is interrelation between confinement and chiral symmetry breaking.

The paper is structured as follows:  in section \ref{Zero}
we describe the physics of the fermionic zero modes and the technical tool -- the hopping matrix --
used to evaluate the determinant. We then explain the general setting of the interactions in section \ref{Setting}. After that we show how the chiral condensate is obtained from the eigenvalue distribution in section \ref{EigenDestribution} and the mass gap is discussed in section \ref{GapWay}. The data sets used and how they were analyzed is explained in section \ref{SectionData}. We end with the physical results in section \ref{Results}, where we show, among other, the Polyakov loop and the chiral condensate's dependence on temperature.

\section{The Zero Mode Zone }\label{Zero}
The term ``dynamical quarks" in the title implies  inclusion of the 
fermionic determinant in the measure for gauge field configurations. 
The main approximation made by us -- similar to what was done in the instanton ensemble --
is that the set of all fermionic states is translated  to the subspace spanned by zero modes.

This determinant can be viewed as a sum of closed fermionic loops with ``hopping amplitudes"
between dyons and antidyons.  Since sectors that are self-dual or anti-self-dual have its eigenvalues protected, then the overlap of $L$ and $L$ dyons or $\bar{L}$ and $\bar{L}$ dyons have to be zero. The resulting form 
of the ``hopping
matrix" is
\begin{eqnarray}
 {\bf \hat T}\equiv \left(\begin{array}{cc}
0&{\bf T}_{ij}\\
-{\bf T}_{ji}&0
\end{array}\right)
\label{T12}
\end{eqnarray}
Each of the entries in ${\bf T}_{ij}$ is a  ``hopping amplitude" for a fermion between
the i-th L-dyon and the j-th $\bar{\rm L}$-antidyon. The diagonal matrix elements are zero,
and therefore a single or many infinitely-separated dyons will have zero determinant and ``veto"
such configurations. However, nonzero non-diagonal hopping matrix elements make 
the determinant nonzero.  

The only modification of the partition function used in this work relative to that in I is the fermionic factor
\be \left( det({\bf \hat T}) \right)^{N_f}
 \ee
Basically, $det({\bf \hat T})$ can be seen as a set of loop diagrams, connecting all L-dyons and antidyons of the ensemble. It can
either be dominated by short loops, including small number (2,..) dyons, to be referred to as a ``molecular regime",
or by very long loops, including finite fraction of the ensemble (``collectivized regime"). The former has unbroken
and the latter broken chiral symmetry. It is the purpose of our simulations to determine,
as a function of the dyon density, the weights of such short and long loops.

We define the individual hopping amplitude as the matrix element of the Dirac operator between different zero-mode eigenfunctions
\begin{eqnarray}
T_{ij} &=& < i | \Dslash | j > \label{eqn_hop}
\end{eqnarray}
where $i$ and $j$ are zero-modes belonging to i-th $L$ and j-th $\bar{L}$ dyons. 
If the gauge field in the Dirac operator is a sum of two solitons, using the equations of motion for
both zero modes, one can reduce the covariant derivative to the ordinary derivative.

Including a mass term, changes the hopping matrix by a constant $m$ times the identity matrix.

\section{The general setting} \label{Setting}

The setup is almost the same as in our paper I \cite{Larsen:2015vaa}, with the difference being the inclusion of 
the fermionic determinant in the zero-modes approximation. This factor creates an additional fermion-induced interaction between the $L$ type dyons.

 The dimensionless holonomy $\nu = v/(2\pi T)$ is related to the expectation value of the Polyakov loop through the ($SU(2)$) relation
\begin{eqnarray}
P &=& \cos  (\pi \nu)
\end{eqnarray}

 We seek to minimize the free energy
\be
f &=& \frac{4 \pi^2}{3}\nu ^2 \bar{\nu}^2 -2n_M\ln\left[\frac{d_\nu e }{n_M}\right] -2n_L\ln\left[\frac{d_{\bar{\nu}} e }{n_L}\right] \nonumber\\
& &+\Delta f \ee
where the first term is the perturbative Gross-Pisarski-Yaffe-Weiss holonomy potential, the next terms
contain semiclassical independent dyon contributions, with
\be d_\nu &=& \Lambda \left( \frac{8\pi ^2}{g^2} \right)^2 e^{-\frac{\nu 8\pi ^2}{g^2}} \nu ^{\frac{8\nu}{3}-1}/(4\pi)
\ee 
and $\Delta f \equiv  -\log (Z_{changed})/\tilde{V_3} $ is defined via the partition function studied numerically
\be Z_{changed}&=& \frac{1}{\tilde V_3^{2(N_L+N_M)}}\int D ^3x \det (G) \exp( -\Delta D_{DD} (x) ) \nonumber \\
& &\times \prod _i \lambda _i ^{N_f}
\ee
The last factor is the fermionic determinant, now written as 
the product of all eigenvalues of the hopping matrix $T_{ij}$. 

Further explanation of $G$ and $\Delta D_{DD}$ can be found in \cite{Larsen:2015vaa}, and we therefore just present their  
expressions here without too many comments.
\begin{eqnarray}
 G &=& \delta _{mn} \delta _{ij} ( 4\pi \nu_m-2\sum _{k\neq i}\frac{e^{-M_D T |x_{i,m}-x_{k,m}|}}{T|x_{i,m}-x_{k,m}|} \\
 & & +2\sum _{k}\frac{e^{-M_D T |x_{i,m}-x_{k,p\neq m}|}}{T|x_{i,m}-x_{k,p\neq m}|} ) \nonumber \\
 & & +2\delta_{mn}\frac{e^{-M_D T |x_{i,m}-x_{j,n}|}}{T|x_{i,m}-x_{j,n}|}-2\delta_{m\neq n}\frac{e^{-M_D T |x_{i,m}-x_{j,n}|}}{T|x_{i,m}-x_{j,n}|} \nonumber
\end{eqnarray}
Dyon 2-point interactions $\Delta D_{DD}$ is a sum over all the different dyon to dyon combinations
\begin{eqnarray}
\Delta D_{DD} &=& \sum _{j >i} \Delta S_{D_i D _j}
\end{eqnarray}
where $\Delta S_{D_i D _j}$ is the correction to the action between dyon i and dyon j. If the two dyons are a dyon and its anti-dyon, we have for distances larger than $x_0$ 
\begin{eqnarray}
\Delta S_{D\bar{D}} &=&  -2\frac{8\pi^2 \nu}{g^2}(\frac{1}{x}-1.632e^{-0.704x})e^{-M_D r T} \nonumber \\
x &=& 2\pi \nu r T
\end{eqnarray}
For the rest of the combinations we have
\begin{eqnarray}
\Delta S_{DD}  &=& \frac{8\pi^2 \nu }{g^2}\left( -e_1e_2\frac{1}{x}+m_1m_2\frac{1}{x}\right)e^{-M_D r T} \nonumber \\
x &=& 2\pi \nu rT
\end{eqnarray}
where the charge is given by table \ref{tab1}. For distances smaller than $x_0$ we have a core between dyon pairs of the types $LL$, $MM$, $\bar{L}\bar{L}$, $\bar{M}\bar{M}$, $L\bar{L}$ and $M\bar{M}$
\begin{eqnarray}
\Delta S_{DD}  &=& \frac{\nu V_0}{1+\exp\left[\sigma T(x-x_0)\right]}\\
x &=& 2\pi \nu rT
\end{eqnarray}
where $x_0$ is the size of the dyons core. In this paper we work with $x_0 =2$, just as in our earlier paper I. It is important to note that for $M$ type dyons one has to use $\nu$ and for $L$ type dyons one has to use $\bar{\nu}=1-\nu$.

\section{Eigenvalue distributions and the chiral condensate}\label{EigenDestribution}
\begin{figure}[h]
\includegraphics[scale=0.48]{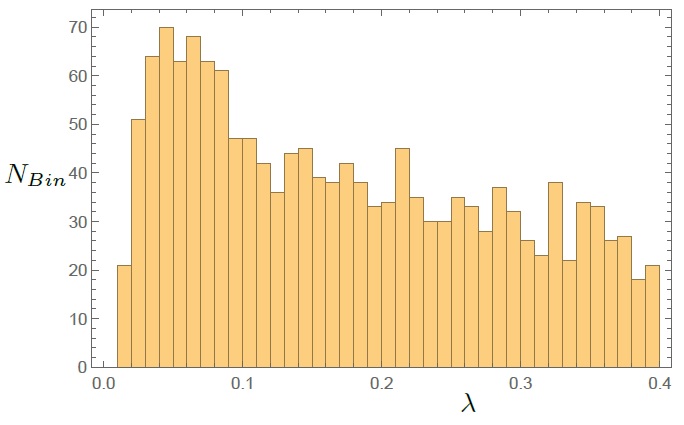}
\caption{Eigenvalue distribution for $n_M=n_L=0.47$, $N_F=2$ massless fermions at $S=7$.}
\label{Eigenvaluesm0Dense}
\end{figure}

\begin{figure}[h]
\includegraphics[scale=0.48]{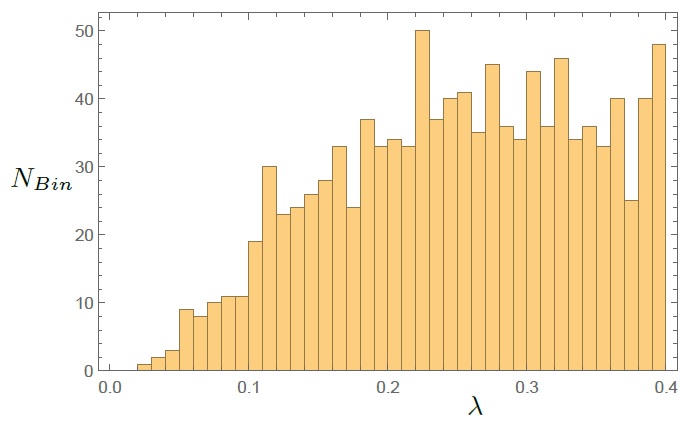}
\caption{Eigenvalue distribution for $n_M=n_L=0.08$, $N_F=2$ massless fermions at $S=7$.}
\label{Eigenvaluesm0}
\end{figure}

\begin{figure}[h]
\includegraphics[scale=0.48]{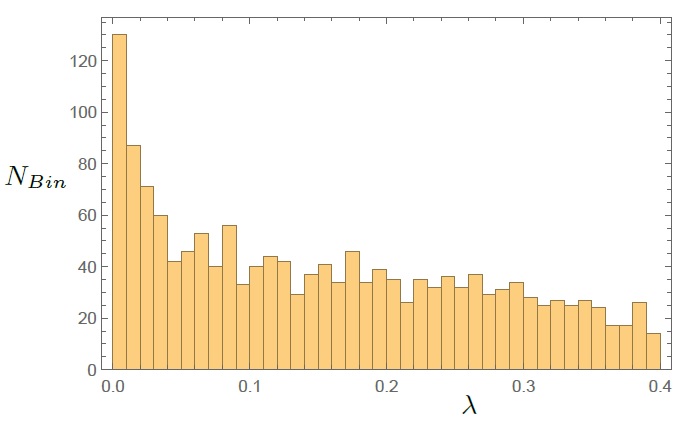}
\caption{Eigenvalue distribution for $n_M=n_L=0.47$, $N_F=2$ $m=0.01$ fermions at $S=7$.}
\label{Eigenvaluesm001Dense}
\end{figure}

\begin{figure}[h]
\includegraphics[scale=0.48]{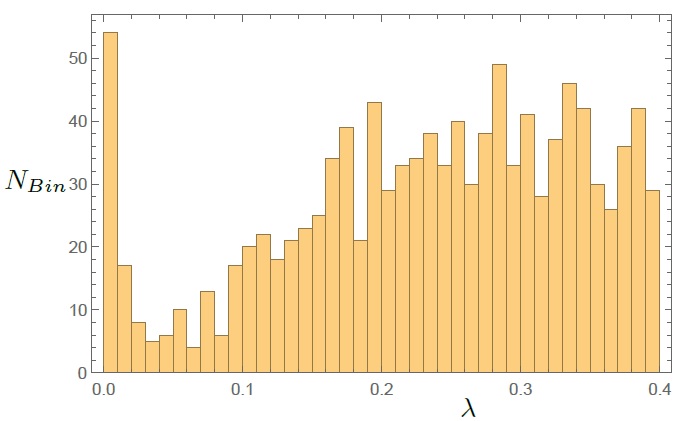}
\caption{Eigenvalue distribution for $n_M=n_L=0.08$, $N_F=2$ $m=0.01$ fermions at $S=7$.}
\label{Eigenvaluesm001}
\end{figure}

The Banks-Casher relation for the chiral condensate tells us that, in the infinite volume limit, the chiral condensate for massless fermions is proportional to
the density of eigenvalues at zero value
\begin{eqnarray}
|<\bar{\psi}\psi >| & = & \pi \rho (\lambda)_{\lambda \to 0,m \to 0,V \to \infty}
\end{eqnarray}
For any system with a finite volume, the typical size of small eigenvalues is of size $1/V$ 
 and the density will always be $0$ at $\lambda =0$ and $m=0$. We see this behavior in our ensemble as seen for zero mass in Fig. \ref{Eigenvaluesm0Dense} and \ref{Eigenvaluesm0}. We also find that a finite mass as in Fig. \ref{Eigenvaluesm001Dense} and \ref{Eigenvaluesm001} has the effect of allowing eigenvalues around zero, and if the mass is large enough, smooth the maximum of the eigenvalue distribution into the region around $\lambda =0$. 
 
 To understand
 finite volume effects on the distribution,  one may study those using  chiral random matrix theory,
 for review see
\cite{Verbaarschot:2009jz}.
 In principle,
 using expressions obtained in this framework one can recover the value of the chiral condensate in the infinite volume case.


We will be determining  the chiral condensate by two different methods:

(i)
The first one is based on the part of the eigenvalue distributions with the smallest $\lambda$. It  
requires an understanding of both the {\em finite volume} and {\em quark mass} effects on the
distribution. This understanding we obtain from analytic random matrix results. We explain this approach in section  \ref{SizeEffects}.


Vanishing of the condensate is used to define the ensemble parameters corresponding
to chiral symmetry breaking transition, $T_{\bar{\psi}\psi}$.

The second  strategy (ii) we will use, is based on the
determination of
  the so called {\em gap width} in the distribution, near $\lambda =0$: we will refer to it as $T_{gap}$. This approach is explained in section \ref{GapWay}.
  
  Ideally, both critical temperatures should coincide, defining the location of the chiral symmetry breaking
  $T_\chi$. 



\subsection{The finite size effects}\label{SizeEffects}
\begin{figure}[h]
\includegraphics[scale=0.46]{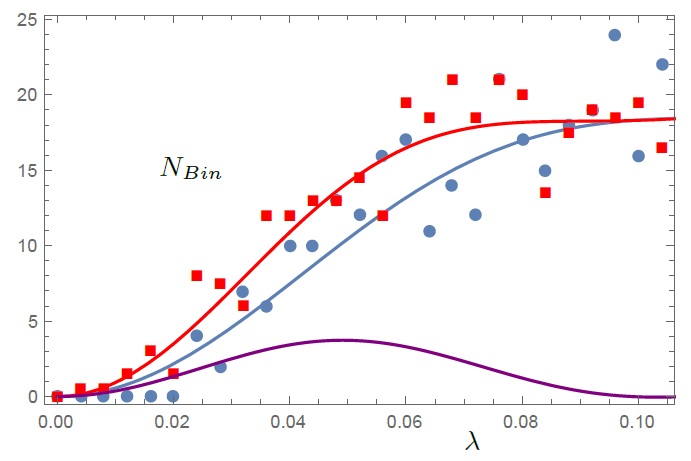}
\caption{(Color Online) The points are the eigenvalue distribution for 64 (blue circles) and 128 (red squares) dyons at $S=8$ and 
density of dyons $n_M=0.33$, $n_L =0.20$, $N_F=2$. The curves are the fit with eq. (\ref{analytic_f})  with  $\Sigma _{2,64}= 1.30 \pm 0.06$ and $\Sigma _{2,128} = 1.28 \pm 0.06$ and the scaling as $\Sigma _{1,64}=0.79 \pm 0.05$ and $\Sigma _{1,128}=0.51\pm 0.04$ for these two cases, respectively. The lower purple line is the difference between the two fits.  Eq. (\ref{SigmaEq}) gives $\Sigma = 0.38\pm 0.13$, while the maximum of the difference between the two curves gives $\Sigma =0.3$ after normalizing the difference (note: This approach of using the maximum of the difference between the two volumes, has not been used to analyze the data, but is simple used here to visualize the effect).}
\label{fig_finite_volume}
\end{figure}

To understand the scaling of the finite volume effects we performed
simulations  for $64$ and $128$ dyons, at the same density.
(The volume of the sphere with $128$ dyons being 2 times larger than the sphere of the $64$ ones.) 
The quark mass in both simulations were set to zero.
The resulting eigenvalue  distributions are shown in Fig. \ref{fig_finite_volume}.

We fit the distribution of the eigenvalues with the form taken from random-matrix theory \cite{Verbaarschot:2009jz} for $SU(2)$  gauge group for massless fermions given by
\begin{eqnarray}
\rho(x) &=& V\Sigma _2[\frac{x}{2}(J_2( x)^2 - J_1( x)J_3( x)) \nonumber
\\ & &+ 
 \frac{1}{2} J_2(x)(1 - \int _0 ^x dt J_2(t)) ]\label{analytic_f}
\end{eqnarray}
where $x=\lambda V \Sigma _1$ and $J_n$ is the Bessel function.  Both the scaling factor $V \Sigma _1$ and the overall factor $V\Sigma_2$ should be proportional to the  value of the chiral condensate $\Sigma$.
In the limit $V\to \infty $ the formula gives $\rho(0)  \propto V\Sigma _2$ as required.

Ideally, the parameter values for two different volumes should agree.
When the fits for different volumes were done, we found that the values for parameter $\Sigma _2$ agree very well indeed.
(This is related to the fact that the height of the distributions at the r.h.s. of Fig. \ref{fig_finite_volume}
do agree.)

Note that the main difference between the two distributions is   a shift to the left for bigger volume.  
This is expected in larger volume clusters of a condensate inside which quark propagation gets larger,
and the eigenvalues smaller. The formula, from random matrix theory, prescribes a particular
``mesoscopic" scaling with  the volume.
However, the fit by this formula produces values of $\Sigma _1$ which are not the same. This indicates that, at least our smaller volume, is not yet 
in the range in which the expected large volume scaling applies.

The physics behind this behavior is as follows: there are basically two components of the ensemble, generating
two different dependencies on the volume. As we already mentioned in the introduction, there is collectivized 
dyons, producing the condensate, and dyon-antidyon pairs. The former
component produces eigenvalue distribution shifting with the volume, while the latter
contribution is volume-independent .

 The existence of two components lead us to construct a value of $\Sigma$ out of all four parameters of the fit given by
\begin{eqnarray}
\Sigma &=& \Sigma _{2}(2\Sigma _{1}^{128}/\Sigma _{1}^{64}-1) \label{SigmaEq}
\end{eqnarray}

\begin{figure}[h]
\includegraphics[scale=0.54]{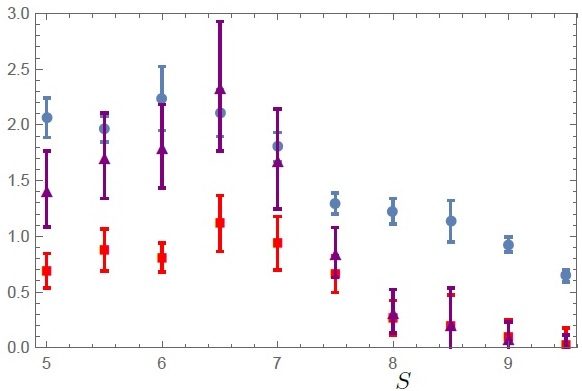}
\caption{(Color Online) $\Sigma_2$ (blue circle), $2\Sigma_1^{128}$/$\Sigma_1^{64}-1$ (red square) and $\Sigma$ (purple triangle) as a function of input action $S=8\pi^2/g^2$ for the results in section \ref{A}. It is observed how the rise in $\Sigma_2$ and $2\Sigma_1^{128}$/$\Sigma_1^{64}-1$ are correlated, while, $2\Sigma_1^{128}/\Sigma_1^{64}-1$ goes to zero for higher $S$ while $\Sigma_2$ does not.}

\label{Sigmacomp}
\end{figure}

In the case of only  almost zero-modes, from the collectivized 
dyons, doubling
 the volume  should double $V \Sigma _1$. In the opposite case of only dyon-antidyon pairs, $V \Sigma _1$ should be unchanged. As can be seen in Fig. \ref{fig_finite_volume} the situation is sometimes in between the two extremes. 
The expression (\ref{SigmaEq}) is an interpolation between the two regimes. 
 This resulting value of $\Sigma$ will be used in the plots to follow, such as showing the temperature dependence
 of the condensate. We show $\Sigma_2$, $2\Sigma_1^{128}$/$\Sigma_1^{64}-1$ and $\Sigma$ for the results in section \ref{A} in Fig. \ref{Sigmacomp}.

As the density increases, it is seen how the scaling becomes closer and closer to that of the volume, as expected from Eq. (\ref{analytic_f}), such that the limit to infinite volume gives the chiral condensate as $\rho (0)$. 


\subsection{The effect of the quark mass}

Nonzero quark mass moderates the distribution of the smaller eigenvalues. Furthermore, for $\lambda <m $
the fermions are effectively decoupled, and thus the distributions should be the same as for a quenched (no 
dynamical quarks) theory. The latter is known to produce a singularity at $\lambda\rightarrow 0$ observed in 
the instanton liquid simulations and on the lattice already in the mid-1990's. 

Our simulations with the mass $0.01$ produce eigenvalue distributions shown in Fig. \ref{Eigenvaluesm001Dense} and \ref{Eigenvaluesm001}. Note  that, in contrast to the zero mass case, one finds  a peak
near zero eigenvalue. Eigenvalues outside of the range of the mass, $\lambda>m$  behave as in the massless case, as can be seen by comparing to Fig. \ref{Eigenvaluesm0Dense} and \ref{Eigenvaluesm0}. In the range of $\lambda = m$ the distribution is smoothed due to the singularity at $\lambda\rightarrow 0$. The same behavior is seen on the lattice \cite{Dick:2015twa}, even when a gap appears. 

\subsection{Gaps of the eigenvalue distribution}\label{GapWay}
At high temperatures --or very dilute dyon ensembles, in our model -- the chiral symmetry
remains unbroken. As it has been shown in multiple lattice simulations, in this case 
the Dirac eigenvalue distribution develops a finite
 gap, between $\lambda=0$ and the point where the eigenvalue distribution starts to rise. 
Vanishing of this gap therefore provides another way of observing the location of the chiral symmetry breaking. Not to confuse it with the critical temperature obtained from the other method,
 we call this temperature for $T_{gap}$.

\begin{figure}[h]
\includegraphics[scale=0.5]{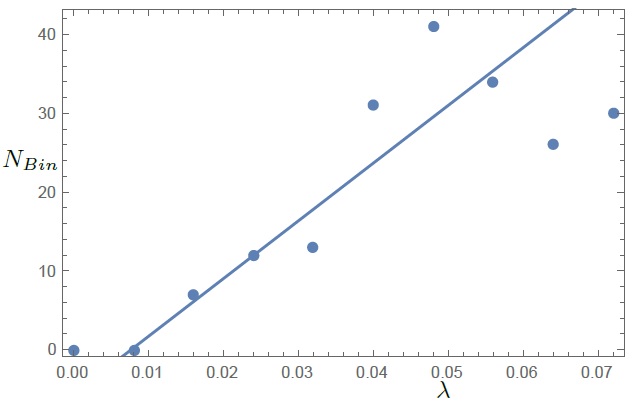}
\caption{The eigenvalue distribution for 64 dyons at $S=7.5$, $\nu=0.434$, $N_F=2$, $n_M=0.43$ and $n_L =0.22$. A straight line has been fitted through point 3 to 6 from the left. The gap size is defined as the cross point with the x-axis. }
\label{fig_gap}
\end{figure}

The procedure used is explained by 
an example shown in Fig. \ref{fig_gap}: 
we fit the distribution by a straight line, and use its intersection with the x-axis as the measure for the gap.

The fact that a gap appears, means that the lowest excitations are not massless.

%



\section{Data and analysis}\label{SectionData}

The setting has already been explained above. 
An ``update cycle" is defined as a sequence of Metropolis updates of all coordinates of  of all dyons. 
Each ``run" consisted of 4000 such ``update cycles", out of which the typical thermal relaxation time was of the order of 500 cycles. The ``useful data"  selected were the mean action values collected for the last 1000 cycles.   

\begin{figure}[h]
\includegraphics[scale=0.5]{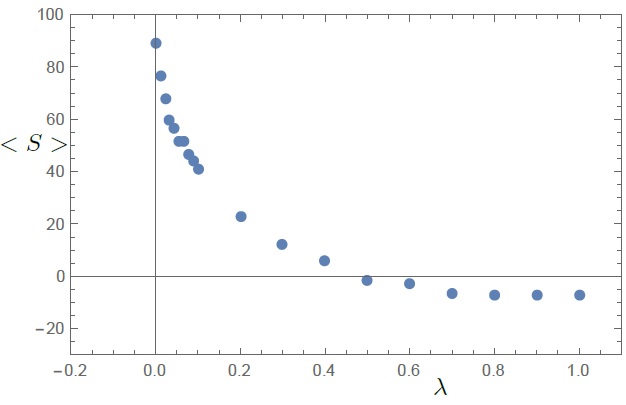}
\caption{A typical example of the expectation values of the action $<S>$ obtained from the simulation as a function of $\lambda$. Contribution to the free energy from the overall constant $F(0)$ is not included. }
\label{fig_lambda}
\end{figure}

The free energy of the model, depending on its parameters, is determined from
  the integrated expectation value of the action $<S(\lambda)>$, following   a standard approach
\begin{eqnarray}
e^{-F(\lambda)} &=& \int D x e^{-\lambda S}\\
F(1) &=& \int _0 ^1 <S(\lambda)> d\lambda +F(0)
\end{eqnarray}
An example of the lambda dependence is illustrated in Fig. \ref{fig_lambda}.
The quick descent in the expectation value of the action at small $\lambda$
required more measurement points in the range $\lambda=0..1$. Therefore we had 
 a step size of $1/90$ until $\lambda=0.1$, while for larger lambda  the step size is increased to 0.1.
 These values, shown in the upper two rows of the Table  \ref{tab2}, constitute 19 runs.

The next three rows of the Table  \ref{tab2} correspond to three parameters of the model
used for free energy minimization. (Those are the value of the holonomy $\nu$, the radius of the system
defining the total dyon density and the number of $M$-type dyons $N_M$.) 
This three-dimensional space was canned systematically, in a lattice form defined by min and max values and
a step defined in the Table. This was done for all values of two remaining ``input parameters",
the Debye mass $M_d$ and classical action $S$. This gives 67200 different combinations. 


\begin{table}[h]
\begin{tabular}{ l | c | c | r  }
  \hline  
 &  Min & Max & Step size  \\   
   \hline 
$\lambda$ & 0 & 0.1 & 1/90  \\
$\lambda$ & 0.1 & 1.0 & 0.1  \\  
$\nu$ & 0.175 & 0.525 & 0.025  \\                    
$r$ & 1.05 & 2.00 & 0.05 \\
$N_M$ & 16 & 26 & 2  \\
$M_d$ & 3 & 6 & 1  \\
$S$ & 5 & 9.5 & 0.5  \\
  \hline  
\end{tabular}
\caption{The input parameters used for the final run.
}
\label{tab2}
\end{table} 

\subsection{Data Analysis}
After the integration over lambda is done, the values of the free energy for each combination of parameters
are determined. The main part of the data analysis is the fit, defining dependence of the free energy
in the 3-dimensional space (of two dyon densities and holonomy) near its minimum. We therefore fit this 
set of data with a 3-dimensional parabola
\begin{eqnarray}
f &=& (v-v_0)M(v-v_0)+f_0
\end{eqnarray}
which has 10 variables. $v$ and $v_0$ are 3D vectors with $v$ containing the variables holonomy $\nu$, radius $r$, and number of M dyons $N_M$ and $v_0$ describing the correction to the point that were the minimum. $M$ is a 3 times 3 matrix with $M=M^T$ containing the coefficients for the fit. 

This expression was fitted to 
 free energy values of $5^3=125$ points from a cube, containing
5 points around the minimum in each direction.  The resulting values of the $10$ parameters fitted
are used as follows:(i)  $v_0$ and its uncertainties give the  values of densities and holonomy
at the minimum, plotted as results below;  (ii) the diagonal component of $M$ in
the holonomy direction was converted into the  value of the Debye mass $M_d$.
An additional requirement of the procedure, to make the ensemble approximately self-consistent,  is  that the Debye mass 
from the fit should be within
$\pm 0.5$ of the used input Debye mass value. 

To obtain the
 chiral properties -- such as the Dirac eigenvalue distributions and its  dependence on dyon number and volume --
we only used the ``dominant" configurations for each action S,
defined as follows. Since $N_M$ is always an integer, we use the value closest to that obtained from the fit.
The eigenvalue distributions is then analyzed as explained in section \ref{SizeEffects}.

\section{Physical Results}\label{Results}

   Accurate gauge-independent
   determination of the hopping matrix element Eq. (\ref{eqn_hop}) is, in general,  not a trivial
 procedure. While zero modes for a single dyon are well known, combining  a pair of  $L$ and $\bar{L}$ dyons
 is not as simple as it is for instantons: the complication is caused by magnetic charges and  the Dirac  
 strings associated with them, transporting singular magnetic flux
to their centers. Ideally those are invisible pure-gauge artifacts, whose direction 
 is irrelevant: but it is not so for simple configurations like the sum ansatz.  ``Combing gauge factors", which appear
   in the zero mode wave function,  complicate the
calculation, although numerically their effect is relatively small: see more  in Appendix A of \cite{Liu:2015jsa}.
 Currently we are working on solving   
 the Dirac equation  for ``streamline" 
configurations defined in \cite{Larsen:2014yya}, but this work is not yet finished. 

As a temporal solution, we use two paramaterizations of the hopping matrix element. 
We perform simulations with both sets. 
 The parameterizations themselves are explained in the Appendix. 
The physical results are, respectively, split up into two sections, one for each choice of $T_{ij}$. Since the overall constant $c'$ is unknown, values of $c'$ have been chosen, such that the transition happens around $S=7.5$. We are actively trying to obtain $c'$ from numerical simulations. While the different $T_{ij}$'s behave similar for large distances, the behavior is different around zero. This also means that the constant $c'$ can be different in the two cases. For these results $c'$ was chosen such that the density of L dyons didn't become too small, while having a smooth Polyakov loop that went to zero in the range of $S=5-10$. 

The plots below have two scales, on their bottom and top. The former one shows the ``instanton action" parameter $S$,
one of the major parameters of the model controlling the diluteness of the ensemble. We also indicate at the top the corresponding temperature, relative to the critical temperature $T_c$, chosen as $S=7.5$. It should be noted that this is a choice, and is done in order to set a scale. The real input is the action $S$ or the coupling constant $g$. The temperature is found from the running coupling constant. 
 \be S(T)={8\pi^2 \over g^2(T)}=b \cdot ln\left({T\over \Lambda}\right),   \,\, b={11\over 3}  N_c -\frac{2}{3}N_F,\ee
This  top temperature scale is 
approximate and should only be used for qualitative comparison to other models and lattice data.

\subsection{Parameterization A for $T_{ij}$}\label{A}

\begin{figure}[h]
\includegraphics[scale=0.50]{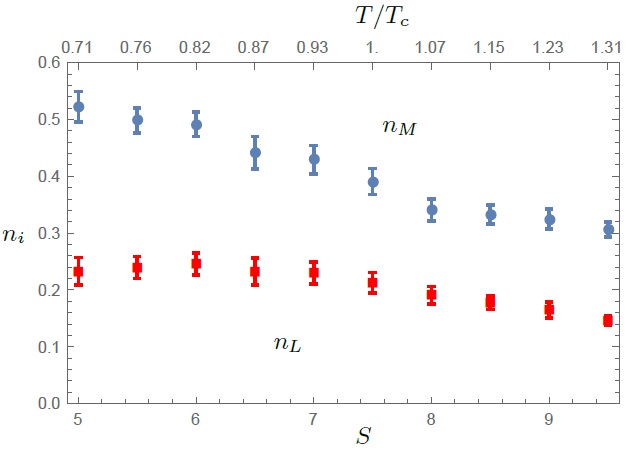}
\caption{(Color online) Parameterization A: The density of the $M$ (blue circles) and $L$ (red squares) dyons as a function of action $S=8\pi^2/g^2$ or temperature $T/T_c$.}
\label{Density_lamb28c2}
\end{figure}

\begin{figure}[h]
\includegraphics[scale=0.5]{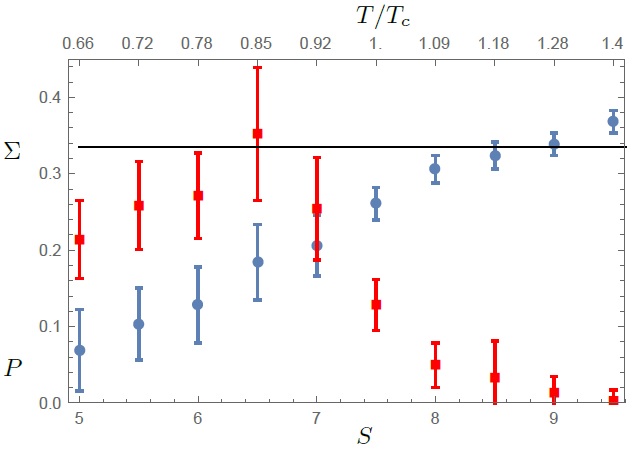}
\caption{(Color online) Parameterization A: The Polyakov loop $P$ (blue circles) and the chiral condensate $\Sigma$ (red squares) as a function of action $S=8\pi^2/g^2$ or temperature $T/T_c$. A clear rise is seen around $S=7.5$ for the chiral condensate. $\Sigma$ is scaled by 0.2. The black constant line corresponds to the upper limit of $\Sigma$ under the assumption that the entire eigenvalue distribution belongs to the almost-zero-mode zone, i.e. the maximum of $\Sigma _2$. }
\label{PandCC_lamb28c2}
\end{figure}

\begin{figure}[h]
\includegraphics[scale=0.5]{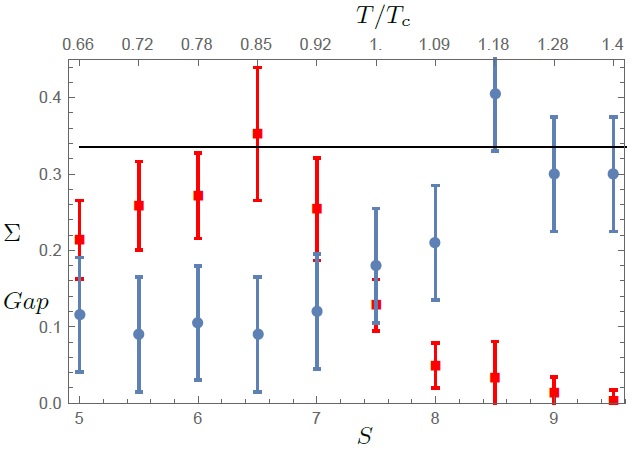}
\caption{(Color online) Parameterization A: The gap scaled up 15 times (blue circles) and the chiral condensate $\Sigma$ (red squares) as a function of action $S=8\pi^2/g^2$ or temperature $T/T_c$. A clear rise/fall is seen around $S=7-7.5$. We get a critical temperature from $S=6.5-8$ for the condensate and $S=6.5-8$ for the gap. $\Sigma$ is scaled by 0.2. The black constant line is defined in the caption of Fig. \ref{PandCC_lamb28c2}.}
\label{gab4}
\end{figure}

\begin{figure}[h]
\includegraphics[scale=0.50]{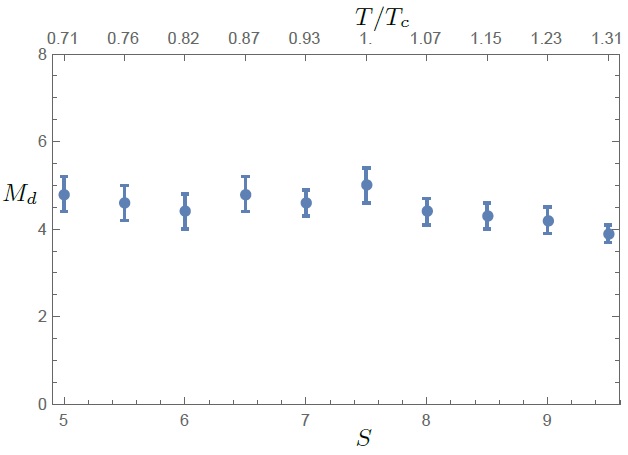}
\caption{Parameterization A: Debye mass $M_d$ as a function of action $S=8\pi^2/g^2$ or temperature $T/T_c$.}
\label{DM_lamb28c2}
\end{figure}

The results in this subsection are for
\begin{eqnarray}
T_{ij} & =& \bar{v} c'\exp{\left(-\sqrt{11.2+(\bar{v}r/2)^2}\right)}
\end{eqnarray}
Minimizing the free energy gives the dominating parameters for a specific action S or Temperature T. This is done for $\Lambda =4$ and $-Log(c')=-2.60$. This gives the holonomy, the density, Fig. \ref{Density_lamb28c2}, and Debye mass, Fig. \ref{DM_lamb28c2}. The dominating configurations have been analyzed using the methods  described in section \ref{EigenDestribution} in order to obtain the chiral condensate, which is shown together with the Polyakov loop in Fig. \ref{PandCC_lamb28c2} and is also compared to the gap in Fig. \ref{gab4}. 

We observe a smooth transition towards zero expectation value of the Polyakov loop $P$ as temperature decreases. We also observe a non-zero value of the Chiral condensate as temperature decreases. This is a more abrupt change, though in some way still smooth. Its inflection point (change of curvature) is found around $S=7.5$, though the transition happens between $S=6.5-8$. Below $S=7$ the results fluctuate around a constant.


  

The chiral symmetry breaking can also be observed through the shrinking of the gap around zero as shown together with the chiral condensate in Fig. \ref{gab4}.  Again, thinking of the inflection points of the two curves, we conclude from it
that the critical temperature for chiral condensate and the gap do coincide within errors,
at the same   $S=6.5-8$ point.

Confinement and chiral symmetry are therefore different phenomena, but are both triggered by the increase in the density of dyons. 

The Debye mass, Fig. \ref{DM_lamb28c2}, as compared to lattice results \cite{Kaczmarek:2005ui},  is seen to be around $66\%$ too large. This could be due to the choice of working with a hard core, or it could signal that the correct value for the size of the core is slightly larger.



\subsection{Parameterization B for $T_{ij}$ }\label{B}

The results in this subsection are for
\begin{eqnarray}
T_{ij} & =& \bar{v} c'\frac{e^{-\bar{v}r/2}}{\sqrt{1+\bar{v}r/2}} \label{T2}
\end{eqnarray}
with $-\log (c') = -0.388$ and $\Lambda = 3.2$. 

\begin{figure}[h]
\includegraphics[scale=0.50]{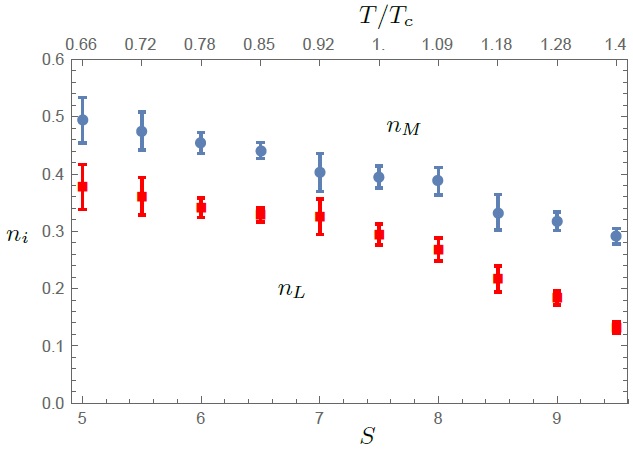}
\caption{(Color online) Parameterization B: The density of the $M$ (blue circles) and $L$ (red squares) dyons as a function of action $S=8\pi^2/g^2$ or temperature $T/T_c$.}
\label{Density_lamb32c1_Norm}
\end{figure}

\begin{figure}[h]
\includegraphics[scale=0.50]{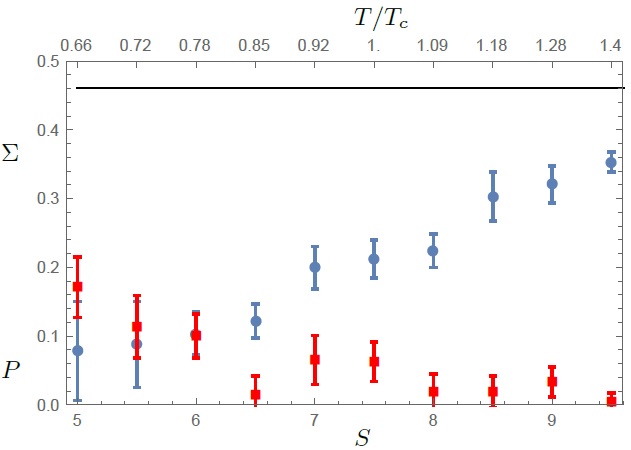}
\caption{(Color online) Parameterization B: The Polyakov loop $P$ (blue circles) and the chiral condensate $\Sigma$ (red squares) as a function of action $S=8\pi^2/g^2$ or temperature $T/T_c$. $\Sigma$ is scaled by 0.1. The black constant line is defined in the caption of Fig. \ref{PandCC_lamb28c2}.}
\label{PandCC_lamb32c1_Norm}
\end{figure}

\begin{figure}[h]
\includegraphics[scale=0.50]{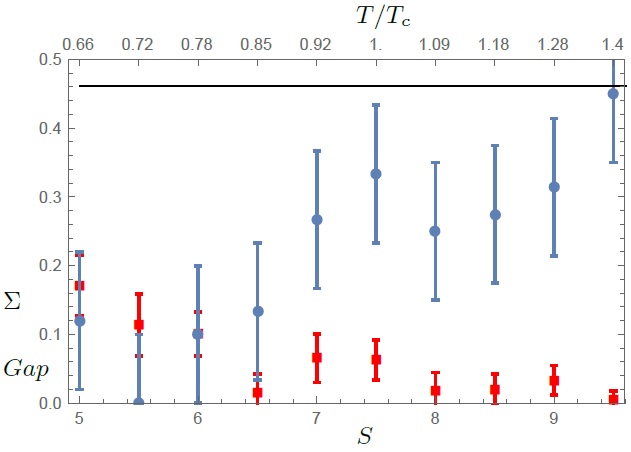}
\caption{(Color online) Parameterization B: The gap scaled up 20 times (blue circles) and the chiral condensate $\Sigma$ (red squares) as a function of action $S=8\pi^2/g^2$ or temperature $T/T_c$. A  fall is seen around $S=7$ for the gap, while it goes close to zero around $S=5-6.5$. At $S=5-6$ the chiral condensate starts to consistently become different from zero. 
It should be noted in this case that $2\Sigma _1^{128}/\Sigma _1^{64}-1$ never becomes larger than $37\% \pm 10\%$. $\Sigma$ is scaled by 0.1. The black constant line is defined in the caption of Fig. \ref{PandCC_lamb28c2}.}
\label{gab32}
\end{figure}

\begin{figure}[h]
\includegraphics[scale=0.50]{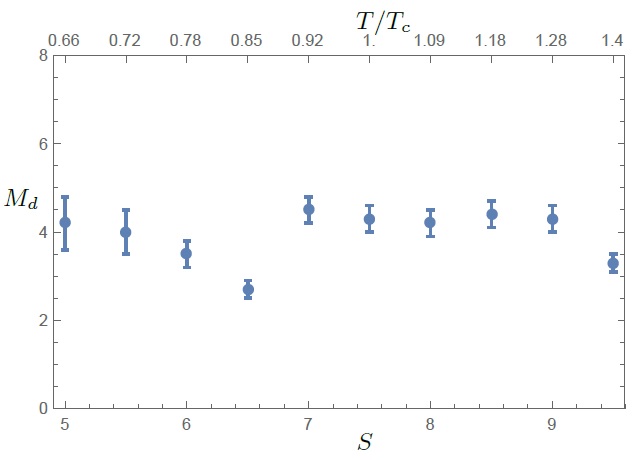}
\caption{Parameterization B: Debye mass $M_d$ as a function of action $S=8\pi^2/g^2$ or temperature $T/T_c$.}
\label{DM_lamb32c1}
\end{figure}

Just as for the other choice of $T_{ij}$ discussed in the previous subsection, we obtain the parameters of density, Fig. \ref{Density_lamb32c1_Norm}, holonomy (Polyakov loop Fig. \ref{PandCC_lamb32c1_Norm}), and Debye mass, Fig. \ref{DM_lamb32c1}, as a function of temperature by minimizing the free energy. The chiral condensate Fig. \ref{PandCC_lamb32c1_Norm} and \ref{gab32}, and gap width Fig. \ref{gab32}, have been obtained from configurations with the parameters obtained by minimizing the free energy. The main difference between the two choices of $T_{ij}$ comes from the behavior around $r=0$. 
The almost exponential behavior as shown in Eq. (\ref{T2}), means that $L$ dyons become more likely at high densities. The other thing is that it is harder to make the different elements in $T_{ij}$ of similar size, which results in a scaling behavior of the chiral condensate that only becomes around $37\% \pm 10\%$ of the volume, and not $100\%$ as with the other choice of $T_{ij}$. This does not mean that the chiral condensate which we show in Fig. \ref{PandCC_lamb32c1_Norm}  does not exist, but it does mean that we need a larger volume in this case to obtain a cleaner result. 
It also means that the overlap between almost-zero-modes and dyon-antidyon pairs was larger.



\section{Conclusion}
We have performed simulations for ensembles of instanton-dyons
 for the setting with two colors $N_c =2$ and two quark flavors $N_f=2$, with variable temperature (coupling constant). We have simulated the partition function for 64 and 128 dyons, calculated the free energy, and derived the values of the
 Polyakov loop,  the chiral condensate and the gaps in the Dirac eigenvalue distributions
 at  the free energy minimum, for each value of the main external parameter $S$ defining the dyon density.
 We also observe gaps in the eigenvalue distribution which goes close to zero in the same interval as the inflection point for the chiral transition.
 
 We find that the required condition for both the chiral symmetry breaking and confinement 
 is 
   basically sufficiently high
  density of the dyons. 
 Furthermore, unlike in the case of pure gauge theory without quarks studied in the previous paper,
 the holonomy dependence on the density is smoother. We don't observe holonomy  vanishing, and also
 the densities of the $M$ and $L$ type dyons does not become equal, even at the lowest $T$ we studied. 
 All of these features make exact determination of $T_c$ difficult and definition-dependent. 
 


It is important to note, that
the repulsive core between the dyons of the same type is essential for these results. For the Polyakov loop expectations value, the core ensures that the holonomy is pushed towards smaller $M$ dyons as density increases, thus making the Polyakov loop expectation value smaller, instead of creating a clump of only $M$ dyons. For the chiral condensate it is important to obtain configurations where the separation from $L$ to $\bar{L}$ dyons are of the same size between the closest dyons, such that the determinant goes from being diagonal dominated between dyon-antidyon pairs to a collective liquid instead. 

While the model itself can definitely be improved -- especially the hopping matrix elements can be defined more accurately --
  the overall mechanism for obtaining confinement and chiral symmetry breaking appears to be very solid, and should not be qualitatively  affected by small changes in the interactions. 
The extensions of the model to other values of $N_c,N_f$ are straightforward,
and we expect to be able to do so in the near future. Another obvious direction of improvement
is larger systems, better statistical accuracy and better control over large volume and quark mass
extrapolations.

\vskip .25cm \textbf{Acknowledgments.} \vskip .2cm 
This work was supported in part by the U.S. Department of Energy, Office of Science under Contract No. DE-FG-88ER40388.

\vspace{0.5cm}

\appendix 
\section{ The hopping amplitudes}

We follow  \cite{Shuryak:2012aa} and
use a simple interpolation formula 
\begin{eqnarray}
T_{ij} & =& c\frac{e^{-\bar{v}r/2}}{\sqrt{1+\bar{v}r/2}}  , \label{Norm}
\end{eqnarray}
where $\bar{v}= 2\pi T-v$. Based on a change of variable it has been found that the constant $c$ should depend on holonomy as $\bar{v}$ which gives
\begin{eqnarray}
T_{ij} & =& \bar{v} c'\frac{e^{-\bar{v}r/2}}{\sqrt{1+\bar{v}r/2}}  ,
\end{eqnarray}

Of course there are many other ways one can choose $T_{ij}$ such that it in the large $r$ limit on a log scale goes as $\bar{v}/2$. We therefore tried to obtain the shape and constant $c'$ from doing first order perturbation theory. 

By doing a first order correction to the energy, it was found that the factor c was dependent on the orientation of the Dirac string, since it was not fixed. The overlap without the gauge transformation was therefore used to understand the shape. The integral done was 
\begin{eqnarray}
\int d^3 x \psi (r_2) \psi(r_1) (\frac{H(r_1)}{2}+K(r_1)) ,
\end{eqnarray}
where $H$ and $K$ are the part of $A_4$ and $A_i$ respectively that only depends on distance and not direction. The shape found to correspond very well to the integral was 
\begin{eqnarray}
T_{ij} & =& \bar{v} c'\exp{\left(-\sqrt{11.2+(\bar{v}r/2)^2}\right)} . \label{Core}
\end{eqnarray}
We will therefore also look into what kind of effect this choice of $T_{ij}$ has.

\begin{figure}[h]
\includegraphics[scale=0.44]{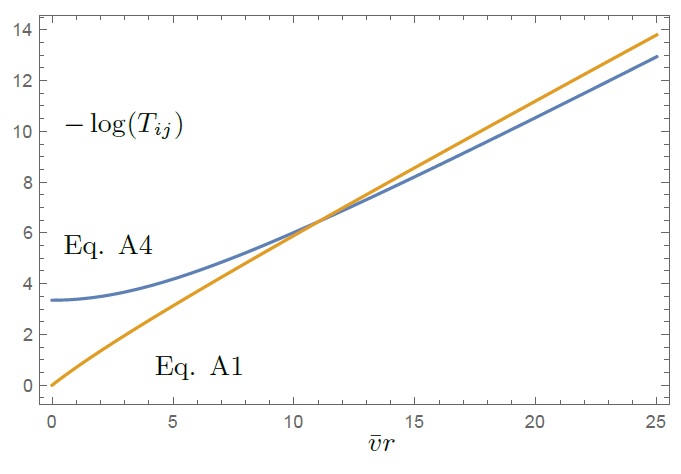}
\caption{The matrix element of the hopping matrix, as $-\log(T_{ij})$
versus the distance, in units of $\bar{v}$. The two different curves corresponds to Eq. \ref{Norm} and \ref{Core}, respectively.}
\label{Comp}
\end{figure}

We compare the two choices in Fig. \ref{Comp}.

\end{document}